\documentclass[12pt]{article}
\pdfoutput=1
\usepackage{amsmath, amssymb, graphicx}
\usepackage[hang,scriptsize,bf]{caption}
\usepackage{cite}
\usepackage{authblk}
\usepackage[margin=1in]{geometry}
\usepackage{color}

\title {Holography, Fractals and the Weyl Anomaly}

\author[1]{Gerald Guralnik\thanks{Deceased, April 26, 2014}}
\author[1]{Zachary Guralnik\thanks{zack@het.brown.edu}}
\author[2]{Cengiz Pehlevan \thanks{cpehlevan@seas.harvard.edu}}
\affil[1]{{\small Physics Department, Brown University, Providence RI.}}
\affil[2]{{\small School of Engineering and Applied Sciences, Harvard University, Cambridge MA.}}

\date{}

\begin{document}
\maketitle
{\hfill \parbox{4cm}{
        Brown-HET-1726 \\
}} 
\vskip 100pt
\begin{abstract}
We study the large source asymptotics of the generating functional in quantum field theory using the holographic renormalization group, and draw comparisons with the asymptotics of the Hopf characteristic function in fractal geometry.  Based on the asymptotic behavior, we find a correspondence relating the Weyl anomaly and the fractal dimension of the Euclidean path integral measure.  We are led to propose an equivalence between the logarithmic ultraviolet divergence of the Shannon entropy of this measure and the integrated Weyl anomaly, reminiscent of a known relation between logarithmic divergences of entanglement entropy and a central charge. It follows that the information dimension associated with the Euclidean path integral measure satisfies a c-theorem.

\end{abstract}
\vskip 100pt
\pagebreak

\section{Introduction}\label{intro}

The large source asymptotics of the generating function in quantum field theory is rarely considered, having no bearing on correlation functions which are defined by the dependence on sources $J$ in the neighborhood of zero.  The thesis of this article is that the large source behavior is of great interest, relating to geometric properties of the Euclidean path integral measure in conformal field theories coupled to gravity. In some instances these properties are fractal-like.
The large $J$ behavior of the Hopf characteristic function of fractals and chaotic invariant sets has been considered in  \cite{FourMeas1,FourMeas2,FourMeas3,ggpHopf,ggpBigChaos}.
In the context of the Hopf characteristic function too, only the small $J$ expansion of the Hopf function is typically of interest, since it contains the moments of the fractal measure.  Aspects of the large $J$ behavior can be determined explicitly for simple fractals and used to compute the fractal dimension and a Lipschitz-H\"older exponent  \cite{ggpHopf,ggpBigChaos}.  With the help of AdS/CFT, or ``holographic'',  duality \cite{Malda,Gubs,Witt}, we will show that there is a correspondence between the Weyl  anomaly \cite{Duff1,Deser1,Duff2,Deser2,Duff} and both the fractal dimension and a Lipschitz-H\"older exponent of the path integral measure.  The exponent in the large $J$ asymptotics which is related to fractal dimension can also be computed explicitly in two dimensional conformal field theory. 

While the large $J$ asymptotics of an interacting quantum field theory is generally difficult to compute, statements regarding asymptotics can be made using  holographic renormalization and  the holographic realization of the Weyl anomaly  \cite{deBoer1,deBoer2,Skenderis,Skend2,Bala,EmparanJohnsonMyers,deHaro,Karch}.
Consider an 
operator ${\Gamma}$
 which is a scalar gauge invariant composite of fundamental fields ${\cal B}_i(x)$ in a  Euclidean conformal field theory for which a holographic dual description exists.
 The generating functional for correlations of ${\Gamma}$ is given by the path integral
\begin{align}
{\cal Z}[{\cal J}] = \int \prod_i{\cal D}{\cal B}_i  e^{-S[{\cal B}] - \int d^d x \sqrt{g}{\cal J}(x){\Gamma}(x)}.
\end{align} 
To address questions of asymptotics, we shall study the behavior of 
\begin{align}
Z(J) \equiv {\cal Z}[J\zeta(x)]
\end{align}
as the $x$-independent-source $J$ becomes large, keeping the arbitrary function $\zeta(x)$ fixed.  The function $Z(J)$ is the generating function for the correlations of the operator
\begin{align}\label{opdef}
{\cal O} \equiv \int d^d x \sqrt{g}\zeta(x) {\Gamma}(x).
\end{align}
Formally, there is a measure over ${\cal O}$ given by 
 \begin{align}\label{PIopmeas}
 d\mu({{\cal O}})= d{{\cal O}}\rho({\cal O})  =  d{{\cal O}} \int \prod_i{\cal D}{\cal B}_i e^{-S[{\cal B}]}\delta\left({{\cal O}} - \int d^d x \sqrt{g} \zeta(x){\Gamma}(x) \right)\, ,
 \end{align}
with respect to which $Z(J)$ is the two sided Laplace transform,
\begin{align}
Z(J) = \int d{\cal O}\rho({\cal O})e^{-J{\cal O}}\, .
\end{align}
In fact, the density $\rho({\cal O})$ is generally ill-defined without introducing a short distance cutoff $\epsilon$, upon which the path integral over fundamental fields depends in a manner described by the renormalization group. In the absence of a smooth $\epsilon\rightarrow 0$ limit of $\rho$,
there is still a well-behaved  limit of $Z(J)$ provided that ${\Gamma}$ is a relevant or marginal operator. The large $J$ asymptotics of $Z(J)$ will be shown to relate to geometric properties of the the limiting measure which are well known in the context of fractals.  The embedding space in which the measure $d\mu({\cal O})$ is defined is one dimensional, as  ${\cal O}$ takes values in one dimensional space.  However the dimension of the measure, known as a fractal dimension, may be non-integer and less than $1$.

To elucidate the connection between large $J$ asymptotics and the geometry of the measure, we first describe this relation in the context of fractal geometry. Consider a fractal set embedded in a $D$ dimensional space parameterized by ${\bf x}$, but having a non-integer fractal dimension $D_f<D$. Such sets do not admit a measure $d\mu({\bf x})$  expressible in terms of a finite density function, $d\mu({\bf x}) \ne d^D{\bf x}\,\rho({\bf x})$. However, as in quantum field theory, one can define a density $\rho_\epsilon({\bf x})$ in the presence of a cutoff $\epsilon$.  This cutoff parameterizes a course graining of the embedding space $\{{\bf x} \}$.    In quantum field theory the cutoff $\epsilon$ corresponds to a course graining of space-time but not, in any obvious way,  a course graining of the embedding space of  quantum fields $\{{\cal O}\}$. A relation between the two will be later be argued to exist.   
For fractals,  the Fourier transform 
\begin{align}
\Psi_\epsilon({\bf J}) \equiv \int d^D{\bf x}\, \rho_\epsilon({\bf x}) e^{-i{\bf J\cdot x}}\, , 
\end{align} 
converges to the Hopf characteristic function $\Psi({\bf J})$ as $\epsilon\rightarrow 0$. Yet, the limiting function $\Psi({\bf J})$, while smooth, is not square integrable and the inverse Fourier transform does not converge.  The $\epsilon\rightarrow 0$ limit of $\rho_\epsilon$ does not exist. 

The absence of a finite density function $\rho(\bf x)$ is related to the existence of a fractal dimension with $D_f < D$. For simplicity, consider a fractal in an embedding space with dimension $D=1$.
At non-zero  cutoff, 
\begin{align} \label{nud}
\int dJ \left|\Psi_\epsilon(J)^2\right| = \int dx \rho_\epsilon^2 \sim \epsilon^{-\nu} 
\end{align}
for some positive $\nu$ \cite{FEDER}.
The parameter $\nu$ measures the divergence of the density function $\rho_\epsilon$, akin to dissipation, along a flow parameterized by $\epsilon$.  
As $\epsilon$ decreases,  $\rho_{\epsilon}$ increases on a shrinking domain of support.
This contraction of the ``phase space'' is responsible for the fractal dimension having a value less than the dimension of the embedding space. The quantity $\alpha \equiv 1-\nu$ is the  Lipschitz-H\"older exponent of a function known as the ``Devil's staircase" \cite{FEDER}, discussed in more detail in section \ref{sec2}.

Although $\Psi(J)$ does not fall off fast enough at large $J$ to be square integrable, 
there exists a minimum real positive number $\gamma$, such that 
\begin{align}\label{gamd}
\int_{-\infty}^{\infty} dJ \left|\Psi(J)^2 J^{-\gamma}\right|
\end{align}
converges.  The quantity $1-\gamma$ is bounded above by the Haussdorf dimension \cite{FourMeas3}, and can itself be taken as a definition of fractal dimension \cite{ggpHopf,ggpBigChaos}.

Explicit examples of the exponents $\nu$ and $\gamma$ will be given for the middle third Cantor set in section \ref{sec2}. 
In section \ref{sec3} we will show that, if a holographic description of a fractal exists,  there is a map relating the fractal dimension to a quantity akin to the Weyl anomaly. 
In section \ref{sec4}, we argue that exponents analogous to $\nu$ and $\gamma$ can  be defined by the large source asymptotics of the generating function in Euclidean conformal field theories having a  holographically dual AdS gravity description.  
The argument is dependent on assumptions of analyticity in $J$, as one is forced to consider the generating function at arbitrarily  large complex values of $J$ and the corresponding dual field in AdS.  Ordinarily, applications of AdS/CFT duality only involve a neighborhood of $J=0$. 

Unlike the exponent $\nu$, $\gamma$ can be computed easily in two dimensional conformal field theory when the source is the background metric, as shown in section \ref{lsrc}.  This confirms the result derived from AdS/CFT arguments.  Although the two dimensional CFT computation is very simple,  the intimate relation between large $J$ asymptotics, the renormalization group and fractal properties is not manifest as it is in the approach based on AdS/CFT duality. 

In the context of conformal field theory, the interpretation of the exponent $\nu$ defined by equation \eqref{nud} differs from the standard case for fractals since the short distance cutoff $\epsilon$ represents a course graining of space-time rather than the embedding space of a field.   
We will argue that a mapping between the short distance cutoff $\epsilon$ and a course graining of the embedding space $\delta$ exists, having the form $\delta(\epsilon) \sim \epsilon^{-d_{\cal O}}$  where $d_{\cal O}$ is the scaling dimension of the field ${\cal O}$ in \eqref{opdef} under Weyl transformations.  Assuming such a map, and writing $\epsilon^{-\nu} = \delta^{-\nu'}$, we shall find that holographic duality implies $\nu'=\gamma$, saturating a bound $\nu'\le \gamma$ characteristic of fractals.  
Note that the exponent $\gamma$, defined by requiring convergence of \eqref{gamd}, is not effected by whether one chooses to look at divergences in terms of $\epsilon$ or $\delta$.
 

Results on the fractal dimension described here  are similar to known results for  entanglement entropy, for reasons discussed in section \ref{sec5}.  In even space-time dimensions, or odd space-time dimensions with a boundary, the coefficient of the $\ln(\epsilon)$  divergent term in entanglement entropy is  proportional to a central charge \cite{CWreplica,Holzhey,Calabrese,Fursaev,Soludukhin,Ryu2,Odds,CasiniFree,Dowker,MyersSinha,Hung,deBoer}.   
In two space-time dimensions, the central charges satisfies the ``c-theorem'' \cite{Zam}, decreasing monotonically under renormalization group flow and providing a measure of the number of degrees of freedom.
Generalizations of this theorem to arbitrary dimensions are based on entanglement entropy  \cite{Latorre,CasiniHuerta1,SolodRicci,KlebRG,Kleb3d,MyersArbDim,CasiniHuerta2,Casini3,Soludatheorem}. 
One definition of fractal dimension, known as the information dimension,  is  the coefficient of the $\ln(\delta)$ divergence  in the Shannon entropy.  
Assuming that the information dimension of the path integral measure over ${\cal O}$ is the same as the fractal dimension determined from the exponent $\gamma$, one concludes that the $\ln(\delta)$  divergence of the Shannon entropy is proportional to the integrated Weyl anomaly.  One then obtains a version of the c-theorem under which the information dimension behaves monotonically under renormalization group flow. 
In a somewhat different context, a connection between entanglement entropy and a fractal dimension has been noted previously in \cite{Astenah}.  In that work, the log divergent term of the entanglement entropy on a fractal entangling sub-region was argued to be related to the fractal dimension and the walk dimension.

\section{Hopf function asymptotics in fractal geometry}\label{sec2}

\begin{figure}[!h]
\center{
	\includegraphics[width=300pt]{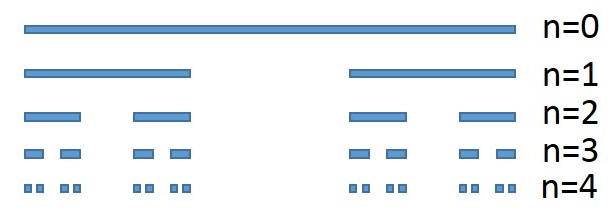}
		\caption{A sequence of operations removing portions of the unit interval, yielding the middle third Cantor set as $n\rightarrow\infty$. For each step in the sequence, one can associate a probability density $\rho_n= \left(\frac{3}{2}\right)^n$ with the remaining parts of the interval. }\label{fig:PictureCantor}}
 \end{figure}

Before studying the large source asymptotics of the generating function in a quantum field theory, it is enlightening  to  consider the asymptotics of the Hopf function for a simple fractal, namely the 
middle third Cantor set.  This set is defined by an infinite sequence of steps in which the middle third of intervals are removed, starting with the unit interval $[0,1]$, as shown in Figure \ref{fig:PictureCantor}.  At the n'th  step, one can define a probability density $\rho_n(x) = \left(\frac{3}{2}\right)^n$ on the remaining intervals, and $\rho_n=0$  where segments have been removed. The measure $d\mu(x) = \rho_n(x) dx$ does not have a smooth $n\rightarrow\infty$ limit of the form $\rho(x) dx$. Yet the $n\rightarrow\infty$ limit of the correlation functions $\int dx x^m \rho_n(x)$ exists. 

\begin{figure}[!h]
\center{
\includegraphics[width=400pt]{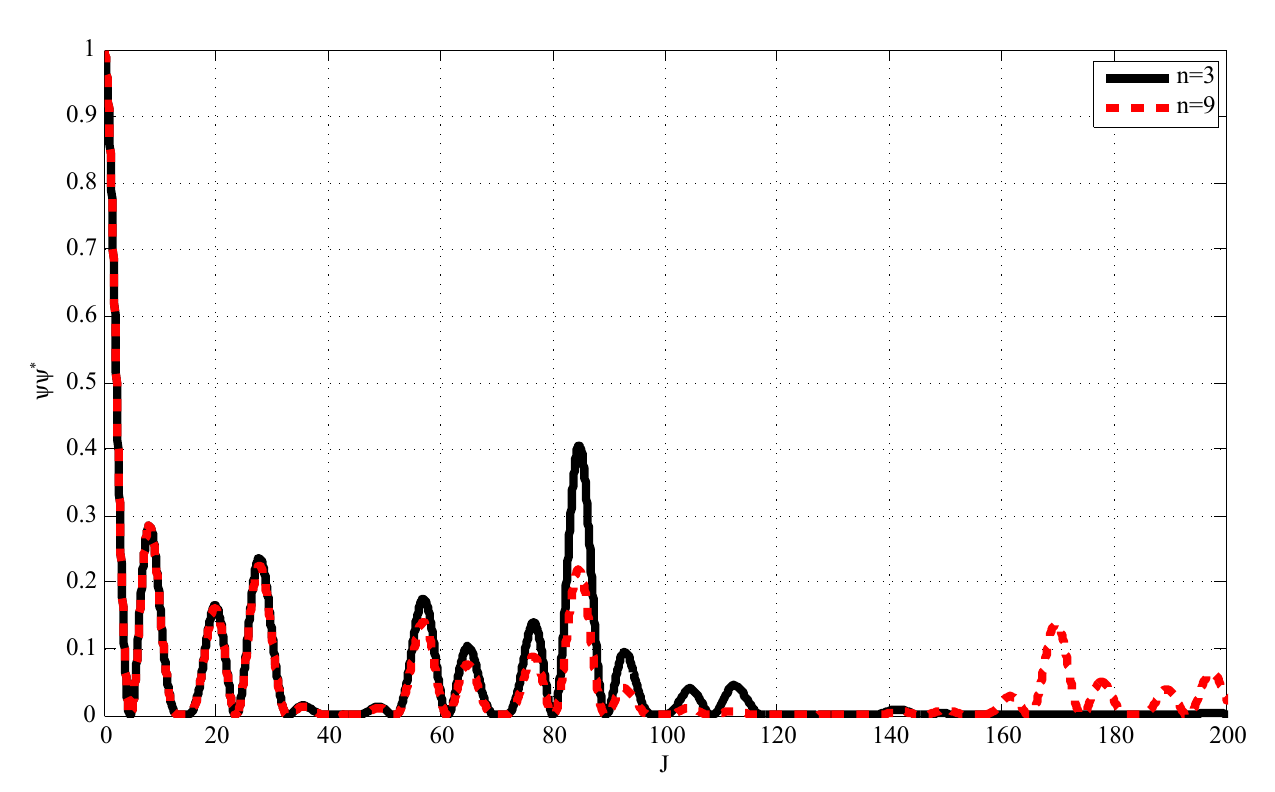}
\caption{The squared amplitude of the Hopf function of the middle third Cantor set, $\Psi\Psi^*(J)$ computed for cutoffs $n=3$ and $n=9$.  The $n=9$ result has already nearly converged to the limiting case over the domain shown, with very small change to be observed by considering higher $n$, whereas the result for n=3 agrees in the smaller $J$ region but is negligiably small for $J >100$.  Despite the absence of a smooth measure in the $n\rightarrow\infty$ limit, the Hopf function converges to a smooth differentiable function.}\label{fig:CantorHopfPlot}
}
\end{figure}

The Hopf function, defined as the Fourier transform of the measure, 
\begin{align}
\Psi_n(J) \equiv \int_0^1 dx \rho_n(x)e^{iJx},
\end{align}
has a smooth limit as the cutoff $n$ is taken to $\infty$ at fixed $J$,
\begin{align} 
\lim_{n\rightarrow\infty}\Psi_n(J)=\Psi(J),
\end{align}
as illustrated in Figure \ref{fig:CantorHopfPlot}.  For any fixed $n$,  $\Psi_n(J)$ falls off rapidly in the large $J$ limit.  However, this is not the case if one first takes $n\rightarrow\infty$, as discussed in \cite{ggpHopf,ggpBigChaos }.  For example, one can show that $\Psi(3^m\pi l )$ is non-zero and independent of $m$ for integer $m$ and $l$.  Upon taking the $n\rightarrow\infty$ limit, the Hopf function ceases to be square integrable, such that its Fourier transform no longer converges.  

Using Parseval's theorem,
\begin{align}\label{divpars}
\int_{-\infty}^\infty dJ \left|\Psi_n(J)^2\right| = \int dx\, \rho_n(x)^2 = \left(\frac{3}{2}\right)^n .
\end{align}
The fact that $\int dx\, \rho_n(x)^2$ diverges with increasing $n$ indicates that $n\rightarrow n+1$ can be thought of as a dissipative map.  As $n$ increases,  the support of the measure becomes increasingly concentrated.  Consequently, the fractal dimension of the Cantor set is less than $1$.  

There exist several definitions of the fractal dimension, often  yielding the same result. One definition, known as the information dimension,  is the coefficient of the logarithmically divergent term in the Shannon entropy,
\begin{align} 
D_{\rm information} = \lim_{\epsilon\rightarrow 0} \frac{-\sum_i P_i \ln(P_i)}{\ln(\epsilon)} \, .
\end{align}
 where the fractal is covered with boxes of length $\epsilon$ and the  ``probability" of being found in the box $i$ is $P_i(\epsilon) \equiv \int_i d\mu(X)$.  
For the middle third Cantor set, the information dimension is easily computed by choosing the cutoff $n$ to be a function of $\epsilon$ such that 
\begin{align}
\epsilon = \frac{1}{3^n}\, .
\end{align}
 Any larger choice for $n$, for a given $\epsilon$,  will yield the same $P_i$. One finds,
\begin{align}\label{dimthird}
D_{\rm information} =  \lim_{\epsilon \rightarrow 0} 
	\frac{\sum_i -P_i \ln(P_i)}{\ln{\epsilon}} = 
\lim_{n\rightarrow\infty}\frac{- 2^n \left(\frac{1}{2}\right)^n \ln{ \left(\frac{1}{2}\right)^n}}{\ln{ \left(\frac{1}{3}\right)^n}} = 
\frac{\ln(2)}{\ln(3)}.
\end{align}

Another definition of the dimension derived from the measure is based on the large $J$ asymptotics of $\Psi(J)$ \cite{ggpHopf,ggpBigChaos}.  Given the minimum value of a parameter $\gamma$ 
such that 
\begin{align}
\int_{-\infty}^{\infty} dJ \left|\Psi(J)^2\right| |J|^{-\gamma}\,,
\end{align} converges,  the fractal dimension can be defined as 
\begin{align}\label{fracd}
D_f =1-\gamma\, .
\end{align}   
In the case of the middle third Cantor set,
\begin{align}
\gamma = 1-\frac{\ln(2)}{\ln(3)}.
\end{align}
This was shown in \cite{ggpHopf,ggpBigChaos}, where the equivalence of the dimension based on large $J$ asymptotics and the information dimension was also shown for other Cantor sets. 
The definition of dimension based on large $J$ asymptotics can also be extended to fractals embedded in (integer) dimensions $D>1$, in which case 
\begin{align}
D_f=D-\gamma\, .
\end{align}

Rather than writing the cutoff measure and Hopf function as $\rho_n, \Psi_n$, we shall henceforward write them as $\rho_\epsilon,\Psi_\epsilon$,  where $\epsilon$ is the course graining, or resolution, of the embedding space given by  $\epsilon = (1/3)^n$ for the middle third Cantor set.  One can then define an exponent $\nu$ by 
\begin{align}\label{fracalph}
\int_{-\infty}^\infty dJ \left|\Psi_\epsilon(J)^2\right| = \int dx \rho_\epsilon(x)^2 \sim \epsilon^{-\nu},
\end{align}
For the middle third Cantor set, \eqref{divpars} gives
$\nu = 1-\frac{\ln(2)}{\ln(3)}$.  Note that in this case $\nu=\gamma$.  

The divergence of \eqref{fracalph} as $\epsilon\rightarrow 0$ is a symptom of the failure of $\Psi(J)$ to fall off sufficiently rapidly at large $J$ for the integral to converge; rapid fall-off at large $J$ only occurs for non-zero $\epsilon$.  The rate at which \eqref{fracalph} diverges as $\epsilon\rightarrow 0$ is related to a Lipschitz-H\"older exponent.  Specifically,
the quantity \begin{align}\label{Holdernu}
\alpha\equiv 1-\nu 
\end{align}
is the a Lipschitz-H\"older exponent of a continuous  non-differentiable function known as the Cantor function or Devil's staircase,
\begin{align}\label{Devil}
f(x) \equiv \int_0^{x} d\mu(x') = \lim_{\epsilon\rightarrow 0} \int_0^x dx'\rho_{\epsilon}(x')  \, .
\end{align}
The Lipschitz-H\"older exponent of a function is defined as 
the maximum value of $\alpha$ such that the bound    
\begin{align}
|f(x+\Delta) -f(x)| \le C\Delta^{\alpha}
\end{align}
is satisfied for some $C$ and all $x$, in the limit of small $\Delta$.  The exponent may also be defined locally (for a given $x$), although we shall be interested in the global version.
For the Cantor function \eqref{Devil}, $f(x+\Delta)-f(x)$ may be evaluated with a  cutoff $\epsilon$ no larger than $\Delta$.  Taking $\epsilon = \Delta$ yields
\begin{align}
f(x+\Delta)-f(x) = \left.\epsilon\rho_\epsilon(x)\right|_{\epsilon=\Delta}  
\end{align}
Therefore, the  expectation value of $|f(x+\Delta)-f(x)|$ is 
\begin{align}\label{expcalc}
\int d\mu(x) \left|f(x+\Delta)-f(x)\right|=  \int dx \rho_\epsilon(x) (\epsilon\rho_\epsilon(x)) = \epsilon^{1-\nu}\, .
\end{align} 
with $\epsilon =\Delta$.
In the small $\Delta$ limit, the expectation value is controlled by the region of $x$ for which the locally defined Lipschitz-H\"older exponent is smallest.  This region in turn determines the global exponent, so that \eqref{expcalc} implies the relation \eqref{Holdernu}.

For a general fractal, one can show that \begin{align}\label{nubound}
\nu \le D-D_b
\end{align}
where $D_b$ is the box counting definition of the fractal dimension. Consider a course graining of the embedding space, such that 
\begin{align}\label{sump2}
\int d^Dx \rho_\epsilon^2 = \epsilon^D \sum_i \left( \frac{P_i}{\epsilon^D} \right)^2 \, .
\end{align}
where the fractal is covered by patches $i$ of volume $\epsilon^D$, and the integral of the fractal measure over each patch is $P_i$.  The number $N$ of such patches scales like $\epsilon^{-D_b}$ where $D_b$ defines the box counting dimension. Subject to the constraint $\sum_i P_i = 1$, \eqref{sump2} is maximized for the case in which all $P_i$ are equal, i.e. maximum Shannon entropy, or $P_i = 1/N$.  Therefore the strongest possible divergence of \eqref{sump2} is given by
\begin{align}
\epsilon^{-D} N^{-1} = \epsilon^{D_b-D}
\end{align} 
implying \eqref{nubound}. 
Provided that the box counting dimension is the same as the dimension defined by $\gamma$, $D_b=D_f$, one has the bound
\begin{align}\label{nuboundJscal}
\nu\le\gamma \, .
\end{align}
It is frequently the case that this bound is saturated, as it is for the Cantor set,  such that the Lipschitz-H\"older exponent is equal to the fractal dimension.

There are computable analogues of the exponents $\nu$ and $\gamma$ for operators in Euclidean conformal field theories with a holographic dual, discussed in sections \ref{sec3} and \ref{sec4}.  
In this context, both $\nu$ and $\gamma$ are dependent on the Weyl anomaly.   However as noted in the introduction, the regularization of the measure does not have the interpretation  as course graining of an embedding space.  Instead $\epsilon$ represents a course graining of space-time. Yet the relation to fractals is more than just an analogy, subject to an assumption that there is an induced course graining of the embedding space of quantum fields which is a function of $\epsilon$.  There is also an assumption that it is possible to analytically continue to large complex sources, outside the usual domain of AdS/CFT duality.  The consequences of these assumptions will be described below.

\section{The fractal hologram: a toy model}\label{sec3}

We shall now take a large leap,  and assume the existence of a fractal having a holographic description resembling AdS/CFT duality.  An immediate consequence of this assumption is a relation between the Lipshitz-H\"older exponent,
fractal dimension and a quantity akin to the Weyl anomaly.  Known AdS/CFT duals will be considered later, but add  complexity which we avoid here for the sake of illustration.

In this supposed  holographic description, one obtains the Hopf function $Z(\bf J)$ of a fractal in an $m$ dimensional embedding space, with  ${\bf J}=J_1\cdots J_m$, from the solution of a  boundary value problem for a diffeomorphism invariant theory with fields ${\bf g}(r) = g_1(r)\cdots g_N(r)$ and $N\ge m$.  The fields $g_I$ satisfy an $r\rightarrow \infty$ boundary condition parameterized by numbers $\hat g_I$, such that the leading large $r$ behavior is,
\begin{align}\label{asympto1}
g_I(r) \sim {\hat g}_I r^{-\beta_I} + \cdots\, ,
\end{align}
along with some fixed initial condition at $r=0$.
Of the coefficients ${\hat g}_I$,  the first $m$ may be regarded as sources,
\begin{align}\label{asympto2}
{\hat g}_I= J_I, \qquad I = 1,\ldots,m,
\end{align} 
while the rest ${\hat g}_{m+1} \cdots {\hat g}_N$ are fixed, corresponding to a parameters defining a particular fractal.
The formal statement of the duality is the relation
\begin{align}\label{basic}
\Psi(i{\bf J})=Z({\bf J}) = e^{-S_{cl}[{\bf g}]}
\end{align}
where $S_{cl}[\bf g]$ is the classical action for the field $\bf g(r)$ on the interval $r=[0,\infty]$, 
satisfying the asymptotics \eqref{asympto1}.   The Hopf function $\Psi({\bf J})$ is defined by the Fourier transform of the fractal measure, whereas $Z({\bf J})$ is the two-sided laplace transform.

Let us further assume that the classical action diverges such that the duality is ill defined without the introduction of a cutoff $\epsilon$, restricting the interval to $r=[0,r_b]$ with $r_b=\frac{1}{\epsilon}$.
Diffeomorphism invariance, or invariance with respect to reparameterizations of $r$, implies that the classical action, viewed as a function of the endpoint $r_b$ and the endpoint boundary conditions ${\bf g}(r_b) = {\bf g}_\epsilon$, for some fixed initial condition, must have no explicit dependence on the endpoint; 
\begin{align}
\frac{\partial S_{cl}({\bf g}_\epsilon,\epsilon)}{\partial \epsilon} =0\, .
\end{align} 
where the partial derivative is taken with the boundary value ${\bf g}_\epsilon$ fixed.
Thus divergences can only arise due to the behavior of the classical solutions ${\bf g}$ at large $r$. 
For our purposes, logarithmic divergences are of primary interest. The regularized classical action can be written as
\begin{align}
S_{cl}({\bf g}_\epsilon) =\frac{1}{2} \ln(\epsilon){\cal A}({\hat g}) + \Gamma(\epsilon,\hat g)
\end{align} 
where $\hat g$ is defined by the asymptotic solution \eqref{asympto1} and $\Gamma$ is finite as $\epsilon\rightarrow 0$, presuming that one has already included potential power law counterterms. The term  ${\cal A}$ is invariant under the re-scaling \begin{align}\label{weylinv} {\hat g}_{I} \rightarrow {\hat g}_{I}\lambda^{\beta_I}\, ,
\end{align}
and is analagous to the Weyl anomaly in AdS/CFT duality \cite{Skenderis,Skend2,Bala,EmparanJohnsonMyers,deHaro}.
Including an explicitly $\epsilon$ dependent counterterm to cancel the log divergence, $S[{\bf g}(r)]\rightarrow S[{\bf g}(r)] -\frac{1}{2} \ln(\epsilon){\cal A}({\bf g}_\epsilon)$,
the equivalence \eqref{basic} is replaced with a renormalized version;
\begin{align}\label{taos}
Z_{\epsilon}({\bf J})
 = e^{ 
-S_{cl}(  {\bf g}_\epsilon   ) 
+\frac{1}{2} {\cal A}( {\bf g}_{\epsilon} )\ln(\epsilon)}\, .
\end{align}

Via arguments described in section \ref{sec2}, geometric properties of the fractal measure, namely dimension and Lipschitz-H\"older exponent, are determined from the dependence of 
\begin{align}\label{complont2}
{\cal F}(\epsilon,\gamma)\equiv \int d^m {\bf J}
\left|Z_\epsilon({\bf J})^2\right| |{\bf J}|^{-\gamma},
\end{align} 
on the cutoff $\epsilon$, as the exponent $\gamma$ is varied.
Strictly speaking, since $Z_\epsilon$ is the two-sided Laplace transform of the measure,  the relation between the cutoff dependance of \eqref{complont2} and fractal properties applies to the case in which the integration is over the imaginary J axes,  or to real integrations if $Z$ is replaced with $\Psi$. The quantity \eqref{complont2} may be computed using \eqref{taos}.  Assuming that holographic duality can be extended to arbitrarily  large complex $J$,  nothing in our arguments will be sensitive to the choice of real vs imaginary axis integration.

The $\epsilon^{-\nu}$ divergence of ${\cal F}(\epsilon,0)$  is directly related to a Lipschitz-H\"older exponent by $\alpha=1-\nu$ only if $\epsilon$ corresponds to a course graining of the embedding space of the fractal.  As will be seen shortly,  $\epsilon$ is non-trivially related to the embedding space resolution $\delta$ in a holographic construction.  

Henceforward, we reserve the notation ${\bf g}(r)$ for bulk fields dual to sources, such that $\hat g= \bf J$, while the remaining bulk fields are written as ${\bf G}(r)$.
For simplicity, we take the classical fields $\bf g$ to have the same large $r$ asymptotic scaling,
\begin{align}
g_{I} \sim J_{I} r^{-\beta}, \qquad I = 1,\ldots,m,
\end{align}
such that in the limit of small $\epsilon$,    
\begin{align}\label{measapprox}
d^m{\bf J} = d^m{\bf g}_\epsilon \epsilon^{-\beta m} + \cdots\, .
\end{align}  
Then \eqref{taos} gives
\begin{align}\label{Zsqd1}
{\cal F} (\epsilon,0)=\int d^m{\bf g}\,\epsilon^{-\beta m} \left|e^{-S_{cl}({\bf g},  {\bf G}_\epsilon) + \frac{1}{2}\ln(\epsilon) {\cal A}({{\bf g}} , {\bf G}_\epsilon) }\right|^2
\end{align}
Note the critical fact that the boundary values ${\bf g}_\epsilon$ have become an integration variable, written just as ${\bf g}$ in \eqref{Zsqd1}, such that $S_{cl}\big({\bf g}, {\bf G}_\epsilon\big)$ may only contain log divergences due to the implicit $\epsilon$ dependence of ${\bf G}_\epsilon$ induced by the classical solutions. If the Weyl anomaly has non-trivial dependence on ${\bf g}$,  the explicit log divergence of the counter-term is no longer fully cancelled by those in $S_{cl}\big({\bf g}, {\bf G}_\epsilon\big)$.  To simplify the discussion, consider the case in which ${\cal A}$ depends solely on ${\bf g}$.
The leading small $\epsilon$ behavior of \eqref{Zsqd1} is then
\begin{align}\label{nug}
{\cal F} (\epsilon,0) & \sim \epsilon^{-\nu}\nonumber \\
 \nu &= \beta m -  {\min}({\cal A})\, ,
\end{align}
where ${\min}{ ({\cal A})}$ is the minimum of the Weyl anomaly, extremized over Weyl equivalence classes of ${\bf g}$ (or $\hat g=\bf J$), with equivalence defined with respect to \eqref{weylinv}.
The integrals over ${\bf g}$ converge via the presumed existence of ${\cal F}(\epsilon, 0)$ and of a measure which is expressible in terms of a probability density function at  non-zero $\epsilon$.  

The fractal dimension is obtained by finding the minimum positive $\gamma$ such that ${\cal F}(\epsilon,\gamma)$ has a finite limit as $\epsilon\rightarrow 0$.  Since in the limit of small $\epsilon$,
\begin{align}\label{srg}
{\cal F}(\epsilon,\gamma) &= \int d^m{\bf g}\,\epsilon^{-\beta(m-\gamma)}\left|e^{-S_{cl}({\bf g},  {\bf G}_\epsilon) + \frac{1}{2}\ln(\epsilon) {\cal A}({{\bf g}} ) }\right|^2
\left| {\bf g} \right|^{-\gamma} \nonumber \\ 
&\sim \epsilon^{-\beta(m-\gamma) + \min{ ({\cal A}) }} \, ,
\end{align} 
the minimum $\gamma$ for which \eqref{srg} is finite is 
\begin{align}\label{epgam}
\gamma = m - \frac{\min{ ( {\cal A} ) }}{\beta}
\, .
\end{align}
  This yields a fractal dimension differing from the embedding dimension $m$, and proportional to the Weyl anomaly:
\begin{align}\label{frarf}
D_f \equiv m-\gamma = \frac{\min{ ( {\cal A} ) }}{\beta}\,.
\end{align}

Note that the exponents $\nu$ of \eqref{nug} and $\gamma$ of \eqref{epgam} are not equal, satisfying $\nu = \beta\gamma$.  
Recall, from the discussion of fractals in section \ref{sec2}, one expects $\nu \le \gamma$ if $\epsilon$ is the resolution of the embedding space, and that this bound is frequently saturated.
Let us define
\begin{align} \label{relnemb}
\delta\equiv \epsilon^\beta\, ,
\end{align}
such that
\begin{align}
{\cal F} (\epsilon,0) & \sim \delta^{-\nu'}\nonumber \\
 \nu' &= m -  \frac{{\min}({\cal A})}{\beta} = \gamma\, .
\end{align}
Thus, if a holographic construction of a fractal exists,  one expects the cutoff $\epsilon$ to be related to the resolution of the embedding space $\delta$ by \eqref{relnemb}.  
Indeed, such a relation arises naturally for AdS/CFT duals discussed in the subsequent section.

Potential obstructions to an explicit construction of a holographic dual of a fractal are described below.  However, even in the absence of an explicit realization,  the arguments above suggest similarities between fractal dimension and the Weyl anomaly in known holographic duals of conformal field theories, which will be explored in more detail in section \ref{sec4}.  

One potential difficulty is that the scaling exponent $\gamma$ is only related to fractal dimension if $Z_\epsilon(J)$ is the Fourier transform of a measure.  Yet the usual formulation of  holography involves the two sided Laplace transform of a measure rather than a Fourier transform. This poses no problem if the $\epsilon$ dependence of ${\cal F}(\epsilon,\gamma)$ of \eqref{complont2} is unchanged in continuing $Z(J)$ to $Z(iJ)$.  

Another complication lies in the application of holographic duality to arbitrarily large $J$.
In order to compute correlation functions,  classical solutions for the bulk fields ${\bf g}$ need only exist  in a small neighborhood of the background  corresponding to ${\bf J}=0$.  The Graham-Lee theorem \cite{GrahamLee}  is an example of an argument guaranteeing the existence of unique solutions \cite{Witt} for sufficiently small variations of the boundary metric in conventional AdS/CFT duality.   We have generally assumed analyticity in the boundary value ${\bf g}_\epsilon$, such that there is no obstruction to considering large $J$ asymptotics. A starting point for an attempt to find a holographic description of a fractal, albeit probably too simplistic, might be an action functional of the form
\begin{align}
S &= \int_{0}^{r_b} dr \left( {\bf\Pi}_{g}\cdot\frac{d {\bf g}}{dr} + {\bf\Pi}_{G}\cdot\frac{d {\bf G}}{dr}- eH({\bf g},{\bf G},{\bf \Pi}_g, {\bf \Pi}_G)\right) \label{sbulk} 
\end{align}
where $H$ a Hamiltonian and $e(r)$ an einbein, or Lagrange multiplier, enforcing $H=0$ or diffeomorphism invariance.
So long as the number of fields $\bf g$ is finite,  there is no obstruction to analyticity.  

Yet another difficulty in finding an explicit example is that the Hopf function of a fractal does not generically have a simple power law scaling behavior as $J\rightarrow \infty$. 
As noted in section \ref{sec2},  there exists an infinite sequence of arbitrarily large real $J$ for which the Hopf function of the middle third Cantor set takes the same value.  It is not clear how such structures can be obtained from a holographic description.  A  crude simplification of the holographic description suggests power law scaling.  Suppose for a moment that the classical fields ${\bf g}$ all have the same $r^{-\beta}$ scaling at large $r$, and let us neglect the fields ${\bf G}$. Consider two different values for the source and for the cutoff which satisfy
$
{\bf J}\epsilon^{\beta} = {\bf J}'{ \epsilon'}^{\beta}\, .
$
The two associated values of the boundary field are approximately equal, since
$
{\bf g}_b = {\bf J}\epsilon^\beta + \cdots \approx {\bf g'}_b
$
where the unwritten terms are sub-leading at small $\epsilon$ and depend on the source.
Then \eqref{taos} implies
\begin{align}
Z_{\epsilon}({\bf J}) \approx Z_{\epsilon'}({\bf J}') e^{ \frac{1}{2}{\cal A}( {\bf\hat g} )\ln\frac{\epsilon}{\epsilon'}} = Z_{\epsilon'}({\bf J}')
e^{\frac{{\cal A}(\hat J) }{2\beta}\ln
\left( \frac{ |{\bf J'}| }{ |{\bf J}|} \right) }\, ,
\end{align}
where because of the Weyl invariance of $\cal A$, one can replace its argument $\hat g$ with the unit vector $\hat J$. 
Taking $\epsilon\rightarrow 0$ and $\epsilon' \rightarrow 0$, 
\begin{align}\label{powr}
Z({\bf J}) \approx Z({\bf J}')
\left( \frac{ |{\bf J'}| }{ |{\bf J}|} \right)^\frac{{\cal A}(\hat J) }{2\beta}
\end{align}
consistent with the power law behavior. 
\begin{align}\label{PL}
Z\sim |{\bf J}|^{\frac{-{\cal A}({ \hat J})}{2\beta}}\, .
\end{align}
The bound on $\gamma$ for convergence of the integral 
${\cal F} \equiv \int d^m{\bf J} |Z({\bf J})|^2 \left| {\bf J} \right|^{-\gamma}
$
at large $\bf J$ is precisely \eqref{epgam}.
Of course, this argument is crude because the behavior \eqref{powr} was obtained only by neglecting the existence of other fields ${\bf G}$ which are considered dual to fixed parameters rather than sources. The pathology at ${\bf J}=0$ which arises here is a symptom of that omission.  

Asymptotic behavior analogous to \eqref{PL} can be shown explicitly when the source is the metric tensor in a two dimensional conformal field theory. In this case the effective gravitational action $\Gamma[g]$, and therefore $Z=\exp(-\Gamma[g])$, is known.
The argument is very simple and can be found in section \ref{lsrc}.  However the relation between the large $J$ asymptotics, the renormalization group  $\epsilon$ dependence and fractal properties is not manifest from $\Gamma[g]$ alone.  These relations are better understood using arguments based on AdS/CFT duality, discussed in the subsequent sections.

\section{Asymptotics of the generating functional from \\AdS/CFT duality}\label{sec4}

The simplified analysis of the previous section can be extended to the large $J$ asymptotics of conformal field theories with a known gravitational holographic dual. 
AdS/CFT duality \cite{Malda,Gubs,Witt} relates the generating function of a conformal Yang-Mills theory in $d$ dimensions to the partition function of a string theory in a space which is asymptotically $AdS_{d+1} \otimes X$,  a direct product of $d+1$ dimensional Anti-deSitter space  with a compact manifold $X$ .  A source ${\cal J}$  coupling to a gauge invariant operator ${\Gamma}$ in the conformal field theory is equivalent to a boundary condition for a dual  field $\phi$ in  Anti-deSitter space. 
The AdS part of the metric can be written in the form, 
\begin{align}\label{HamGauge}
ds^2_{(d+1)} = 
dr^2 + g_{ij}(x,r) dx^i dx^j\, , 
\end{align}
with boundary value
\begin{align}\label{CFTmet}
\lim_{r\rightarrow \infty} e^{-2r/L} g_{ij}(x,r) = \hat g_{ij}(x)
\end{align}
corresponding to the metric of the conformal field theory.  Henceforth, we take units such that the radius of AdS space is  $L=1$. 
The large $N$, strong 't Hooft coupling limit of the conformal field theory generating functions can be obtained from the classical gravitational theory in Anti-deSitter space:
\begin{align}\label{ZED}
{\cal Z}[{{\cal J}}]\equiv
\left\langle \exp\left( \int d^d x\,\sqrt{ \hat g}{{\cal J}_a(x)} {\Gamma_a(x)}\right)  \right\rangle
 = \exp\left(-S^{cl}_{GR}\right)\,  ,
\end{align} 
where  $S^{cl}_{GR}$ is the classical action of the $d+1$ dimensional bulk theory including gravity and other fields,  
with the leading large $r$ behavior of the bulk fields $\phi_a$ dual to $\Gamma_a$ is given by\begin{align}\label{bbhv}
{\phi}_a(x,r) = {{\cal J}_a}(x)e^{-\beta r}\ +\cdots\,\, .
\end{align}
To simplify the following discussion, the only bulk fields we consider are the metric $g_{ij}(x,r)$ and a scalar field $\phi(x,r)$.
An operator $\Gamma$ in the conformal field theory dual to a scalar field $\phi$ in AdS has conformal scaling dimension $\lambda = d-\beta$. 

The classical action $S^{cl}_{GR}$ diverges if the integral over the AdS space time is taken all the way to the boundary at $r=\infty$. 
The gravitational part of the regularized action is,
\begin{align}
S_{regularized} = \frac{1}{16\pi G_N^{(d+1)}}\left[\int_{\Sigma} dr \,d^d x\sqrt{ g^{(d+1)}}(R+ 2\Lambda ) + \int_{\partial\Sigma} d^d x \sqrt{ h^{(d)}}K\right]. 
\end{align}
where the boundary $\partial\Sigma$ of the bulk space-time $\Sigma$ lies at finite $r=r_b$.  For the AdS metric in the form given by \eqref{HamGauge} and \eqref{CFTmet}, the boundary is related to a short distance cutoff $\epsilon$ in the conformal field theory by $e^{-r_b}=\epsilon$.  The induced metric on the boundary is $h^{(d)}_{ij}$, equal to the boundary value of $g_{ij}$ defined by \eqref{HamGauge},  while $K$ is the Gibbons-Hawking-York term \cite{HawkGib,York}. The latter is built from the trace of the extrinsic curvature of the boundary, and is necessary for a well posed variational principle in the presence of a boundary.  Divergences of the classical action as $\epsilon\rightarrow 0$ are proportional to covariant local functionals of the CFT metric $\hat g$ \cite{Skenderis}.
Remarkably, those divergences which are negative powers of $\epsilon$ may be cancelled by counterterms having no explicit $\epsilon$ dependence, involving 
covariant local functionals of the induced boundary metric $g_{ij}(x,r_b) \equiv g^{\epsilon}_{ij}(x)$   \cite{Bala,EmparanJohnsonMyers}. Such terms have implicit dependence on $\epsilon$, diverging only because of the behavior of the AdS metric at large $r$. For even $d$ there may be additional divergences proportional to $\ln(\epsilon)$,  which are canceled only by the addition of boundary counter-terms depending explicitly on $\epsilon$.  The counter-term action has the form
\begin{align}\label{counterterm}
S_{counterterm} = \int_{\partial\Sigma}d^d x\sqrt{{h^{(d)}}}\left( {\Omega} - \frac{1}{2}\ln(\epsilon){\cal A}\right)
\end{align}
where ${\Omega}$ and $\cal A$ are local functions of the bulk fields and induced metric evaluated at the boundary. 
The term ${\cal A}$ corresponds to the the Weyl, or conformal, anomaly \cite{Skenderis,Skend2,Bala,EmparanJohnsonMyers,deHaro,Karch}.  

An important feature of the regularized classical action $S^{cl}_{regularized}$, expressed as a function of boundary data, is that it has no explicit dependence on $\epsilon$. 
In a Hamiltonian formulation of gravity used in the context of the holographic renormalization group in \cite{deBoer1,deBoer2}, with the metric expressed in the form \eqref{HamGauge}, the Hamiltonian describes evolution in the $r$ direction.  Diffeomorphism invariance implies that the on-shell Hamiltonian vanishes, such that the Hamilton-Jacobi equation gives
\begin{align}\label{HJac}
\frac{\partial}{\partial \epsilon} S^{cl}_{regularized}[ \phi_\epsilon,g^{ij}_\epsilon,\epsilon] =H=0,
\end{align} 
where the partial derivative  is taken with the boundary fields $\phi_\epsilon$ and $g^{ij}_\epsilon$ held fixed. 
However, due to the Weyl anomaly, the renormalized classical action has explicit $\epsilon$ dependence;
\begin{align}
S^{cl}_{renormalized} = S^{cl}_{regularized}[ \phi_\epsilon,g^{ij}_\epsilon ] + S_{counterterm}[\phi_\epsilon,g^{ij}_\epsilon,\epsilon]
\end{align}
It will be convenient to write the renormalized action as
\begin{align}
S_{renormalized} = \tilde S -\int_{\partial\Sigma} d^dx\sqrt{h^{(d)}}\frac{1}{2} \ln(\epsilon){\cal A}\, ,
\end{align}
where 
\begin{align}
\tilde S \equiv S_{regularized} + \int_{\partial\Sigma}d^d x\sqrt{{h^{(d)}}} {\Omega}
\end{align}
On shell, $\tilde S= \tilde S_{cl}[\phi_\epsilon,g^{ij}_\epsilon]$, having no explicit $\epsilon$ dependence but diverging logarithmically due to the behavior of classical solutions at large $r$.  In the following we use the equivalence
\begin{align}
{\cal Z}_\epsilon({\cal J}) &= e^{ 
	-\tilde S_{cl}[ \phi_\epsilon,g^{ij}_\epsilon ] 
	+ \frac{1}{2}\ln(\epsilon) A[\phi_\epsilon,g^{ij}_\epsilon] 
	} \nonumber\\
A&\equiv \int_{\partial\Sigma} d^dx\sqrt{h^{(d)}} {\cal A}
\end{align}
to analyze the large ${\cal J}$ asymptotics of ${\cal Z}$.


The embedding dimension of the measure over ${\Gamma}(x)$ in \eqref{ZED} is infinite, as there are an infinite number of space-time points.  To simplify matters, we shall apply holography to study the large $J$ asymptotics of 
\begin{align}\label{simpp}
 Z(J)\equiv {\cal Z}[{{\cal J}} = J\zeta(x)]\, , 
\end{align}
where $J$ is independent of $x$ and $\zeta$ is a fixed arbitrary function of $x$. $Z(J)$ is the the generating function for the operator 
\begin{align}
{\cal O}\equiv\int d^d x \sqrt{ g^{(d)}} \zeta(x) {\Gamma}(x)\, ,
\end{align}
such that the measure over ${\cal O}$ has embedding dimension $1$.
For finite cutoff, the measure can be written as 
\begin{align} 
d\mu({\cal O}) = \rho_{\epsilon}({\cal O})d{\cal O}\, ,
\end{align}
the density $\rho_{\epsilon}$, with path integral definition given formally by \eqref{PIopmeas}, is  
the inverse Laplace transform \footnote{The inverse of the one-sided Laplace transform frequently requires an integration over the axis $Re(J)=c$, for sufficiently large positive $c$, so as to be within the region of convergence of the Laplace transform.  In the present instance, one may take $c=0$.}
\begin{align}
\rho_{\epsilon}({\cal O}) =\frac{1}{2\pi i}\int_{-i\infty}^{+i\infty} dJ\,  e^{J{\cal O}} Z_{\epsilon}(J)\, .
\end{align}
Conditions for the existence of this integral are the same as that for the Fourier transform of $Z(iJ_I)$ with respect to the (real) variable $J_I$.  We presume that the sufficient condition of square integrability with respect to $J_I$ is generically met at finite cutoff.

Motivated by the discussion in section \ref{sec3}, we seek the cutoff dependence of 
\begin{align}\label{sorg2}
{\cal F}(\epsilon,\gamma) \equiv 
\int d{ J} \left|Z_\epsilon({ J})^2  {J}^{-\gamma}\right|\, .
\end{align}
The integration over the imaginary axis has direct bearing on the geometry of the measure, whereas one normally only considers real $J$ in the context of AdS/CFT duality. 
Even though UV counterterms render the effective action finite, such that $\lim_{\epsilon\rightarrow 0} Z_{\epsilon}(J)$ is finite,   ultraviolet divergences may still persist in the integral \eqref{sorg2}.  Analogous persistent divergences were shown explicitly for the Hopf function of a Cantor set in section \ref{sec2}, and essentially the same phenomenon is at play here.
 Assuming no dependence on any dimensionful parameters in the function $\zeta(x)$ of \eqref{simpp}, conformal invariance forces a divergence of \eqref{sorg2}, if it exists, to have the form
\begin{align}\label{pseudotrivial}
{\cal F}(\epsilon,\gamma) \sim \epsilon^{-(d-\lambda)(1-\gamma)}.
\end{align}
Using holographic duality  we shall find this divergence,  modified  non-trivially in cases for which the source contributes to the Weyl anomaly. 
Nothing in the holographic arguments suggests that the  exponent in the divergence is changed by taking the integration in \eqref{sorg2} over the imaginary axis, in which case the results yield information about the geometry of the measure, in the form of a fractal dimension and Lipshitz-H\"older exponent.  However it is presumed, perhaps brazenly, that holography extends to arbitrarily large complex $J$.

A quantity analogous to a Lipshitz-H\"older exponent follows from consideration of ${\cal F}(\epsilon,0)$.
Applying holographic duality to compute $Z_{\epsilon}(J)$ yields,
\begin{align}\label{holoref_in}
{\cal F}(\epsilon,0) = 
 \int d J \left| \exp\left(-\tilde S_{cl}[ \phi_\epsilon,g^{ij}_\epsilon] + \frac{1}{2}\ln(\epsilon){ A}[\phi_\epsilon,g^{ij}_\epsilon]\right)\right|^2 \, ,
 \end{align}
 with 
\begin{align}
\phi_\epsilon &\equiv \phi_{cl}(x,r_b), \,\,\,\,\,\,\, \phi_{cl}(x,r) = J\zeta(x) e^{-r(d-\lambda)} + \cdots \label{asympexp}   \\
g^{ij}_{\epsilon} &\equiv g_{cl}^{ij}(x,r_b), \,\,\,\,\,\,\, g_{cl}^{ij}(x,r) = e^{-2r}\hat g^{ij}(x) + \cdots\, , \\
&e^{-r_b} =\epsilon\, .
\end{align}
Defining 
\begin{align}
\xi \equiv J \epsilon^{d-\lambda}
\end{align}
equation \eqref{holoref_in} becomes,
\begin{align}
{\cal F}(\epsilon,0) \label{holodummy}
\approx\int d \xi\, \epsilon^{\lambda-d} 
\left| 
	\exp\left( 
			-\tilde S_{cl} \left[ \xi\zeta(x),g^{ij}_\epsilon(x) \right]
			+\frac{1}{2}\ln(\epsilon){ A}\left[ \xi\zeta(x),g^{ij}_\epsilon(x) \right]
		\right)
\right|^2.
\end{align}
The relation is approximate due to the absence of subleading terms at small $\epsilon$. 
Note that any $\epsilon$ dependence of  \eqref{holodummy} is either explicit or lies in the dependence of the classical solution for $g_{ij}$ on $r$, but does not reside in the dependence of the classical solution for $\phi$ on $r$. The boundary value of the field $\phi$, dual to the source $J$, is now an integration variable for which the $\epsilon$ dependence has been pushed into the measure $dJ = d\xi \epsilon^{\lambda-d}$. 

We assume for the moment that the Weyl anomaly does not depend on $\phi$, as it would for example in a dilaton coupled CFT.
Then \eqref{holodummy} becomes, 
\begin{align}\label{shmerg}
{\cal F}(\epsilon,0)  \approx
\int d \xi\, \epsilon^{\lambda-d} 
\left| 
\exp\left(  
	-\tilde S_{cl}\left[ \xi\zeta(x),g^{ij}_\epsilon(x) \right]
	+\frac{1}{2} \ln(\epsilon){ A}[\hat g_{ij}(x)] 
	 \right)
\right|^2.
\end{align}
where we have replaced $g^{ij}_\epsilon$ with $\hat g_{ij}$ in the argument of ${ A}$ due to its Weyl invariance.
The anomalous term cancels the log divergence of $\tilde S$ arising because of the dependence of $g^{ij}_{\epsilon}(x)$ on $\epsilon$ induced by the classical solution, giving 
\begin{align}
{\cal F}(\epsilon,0) &\sim \epsilon^{-\nu} \nonumber \\
\nu &=  d- \lambda\, ,
\end{align}
One can not immediately relate the exponent $\nu$ to an analogue of a Lipshitz-H\"older exponent since $\epsilon$ corresponds to a space-time cutoff rather than a resolution $\delta$ in an embedding space for the field ${\cal O}$ to which the source couples.  If the regularized measure over ${\cal O}$ can also be interpreted as a course graining of the embedding space, then there is a relation between $\epsilon$ and $\delta$. 
 Comparing the weights of ${\cal O}$ and $\epsilon$ under a Weyl transformation suggests $\delta \sim \epsilon^{d-\lambda}$, with ${\cal O}$ assumed to be a relevant operator; $d-\lambda>0$.  Then
\begin{align}\label{fstnu}
{\cal F}(\epsilon,0) &\sim \delta^{-\nu'} \nonumber \\
\nu' &=  1\, .
\end{align}
If the exponents $\nu'$ and $\gamma$ are indeed related to  a Lipshitz-H\"older exponent and fractal dimension, we expect to find $\gamma =\nu'$  below.
As noted above, the correspondence between these exponents and fractal geometry is more than just an analogy, provided that the cutoff dependence is unchanged by taking the $J$ integration along the imaginary axis.

To compute the analogue of fractal dimension, we consider the quantity,  
\begin{align}\label{derp}
{\cal F}(\epsilon,\gamma)&\equiv\int dJ \left|Z(J)^2 J^{-\gamma}\right|   \\
&\approx \int d \xi\, \epsilon^{\lambda-d} 
\left| 
\exp\left(
	-\tilde S_{cl}\left[ \xi\zeta(x),g^{ij}_\epsilon(x) \right]
	+\frac{1}{2} \ln(\epsilon){ A}[\hat g_{ij}(x)] 
	 \right)
\right|^2\left|\xi\right|^{-\gamma}\epsilon^{-\gamma(\lambda-d)}\nonumber \, .
\end{align} 
As in \eqref{shmerg}, the anomalous term cancels the log divergence of $\tilde S_{cl}$ due to the dependence of $g^{ij}_{\epsilon}(x)$ on $\epsilon$. Therefore one finds the cutoff dependence,
\begin{align}
{\cal F}(\epsilon,\gamma) \sim \epsilon^{(\lambda -d)(1-\gamma)}\, .
\end{align}
The analogue of fractal dimension is obtained by finding the minimum value of the positive parameter $\gamma$, such that\eqref{derp} converges in the $\epsilon\rightarrow 0$ limit.  Thus $\gamma=1$  for  a relevant operator with $\lambda < d$.  In light of \eqref{fstnu}, the bound $\nu' \le \gamma$ characteristic of fractals is saturated. For a marginal operator with $\lambda=d$, $\gamma=0$.  The definition of fractal dimension based on the exponent $\gamma$ is $D_f = 1-\gamma$, so that
\begin{align}
D_f &= 0: \,\,\, {\rm relevant}\nonumber \\
D_f &= 1:\,\,\, {\rm marginal} \label{marginal}.
\end{align}
Although integer, the fractal dimension is still non-trivial for relevant operators, being less than the embedding dimension.  
Non-integer results for the fractal dimension can arise only if the Weyl anomaly depends upon the source. 

Let us therefore consider a source parameterizing a deformation of the CFT metric, such that
\begin{align}\label{jilg}
g^{ij}(x,r) = \left(\hat g^{ij}(x) + J\sigma^{ij}(x)\right)e^{-2r} + \cdots\, ,
\end{align}
where $\sigma^{ij}$ is symmetric.  AdS/CFT duality relates the corresponding bulk partition function to a  
generating function for stress tensor correlations of the CFT with metric $\hat g$, 
\begin{align} 
Z(J) =  \exp{\left(J\int d^d x \, \sigma^{ij}\frac{\delta}{\delta {g'}^{ij}} \right)} {\cal Z}[ g']_{g'=\hat g} \, .
\end{align}
Defining 
\begin{align}
{\cal G}\equiv J\epsilon^2
\end{align}
duality implies
\begin{align}\label{zergl}
{\cal F}(\epsilon,0) &\equiv \int dJ |Z_\epsilon(J)|^2  \\ 
&= \int d{\cal G}\epsilon^{-2}
\left|
\exp\left(
-\tilde S_{cl}[ g^{ij}_\epsilon] ) + \frac{1}{2}\ln(\epsilon){ A}[g^{ij}_\epsilon]
\right)
\right|^2 \,,
\label{shmu}
\end{align}
where the boundary value of the metric is
\begin{align}
g^{ij}_\epsilon =  {\cal G}\sigma^{ij}+\hat g^{ij}\epsilon^2 +\cdots \, .
\end{align}
	
We have thus far ignored the fact that there are real $J$ values outside a neighborhood of $J=0$ for which the boundary metric will be singular. 
If $\sigma_{ij}$ is not positive definite, then  $\hat g^{ij} + J\sigma^{ij}$ will only be positive definite within a strip containing $J=0$: $J_-<J<J_+$. 
If $\sigma_{ij}$ is positive definite, then one requires $J>J_-$ for some negative $J_-$.  This is consistent with the understanding of $Z(J)$ here as a one-sided Laplace transform, with measure restricted to positive ${\cal O}=\int d^d x \sqrt{\hat g} \sigma^{ij}T_{ij} $.
Strictly speaking, obtaining results related to fractal properties requires consideration of $Z(J)$ for  large imaginary $J$, evaluating the integral \eqref{zergl} along an axis at constant real J (e.g $Re(J)=0$), and presumably evading singularities.   For now we shall assume $\sigma^{ij}$ to be positive definite and, viewed as a metric itself, topologically equivalent to $\hat g^{ij}$. The case in which $\sigma^{ij}$ is non-vanishing over a subset of the CFT space-time is interesting but will not be considered here. 

To leading order in small $\epsilon$, the large $J$ (or large ${\cal G}$) contribution to the integral \eqref{shmu} is,
\begin{align}
{\cal F}(\epsilon,0) &\approx 
\int d{\cal G}\epsilon^{-2}
\left|
\exp\left(
-\tilde S_{cl}[ {\cal G}\sigma^{ij} ] + \frac{1}{2}\ln(\epsilon){ A}[ {\cal G}\sigma^{ij}]
\right)
\right|^2\nonumber \\
&= \int d{\cal G}\epsilon^{-2 + { A}[\sigma^{ij}]}
\left|
\exp\left(
-\tilde S_{cl}[ {\cal G}\sigma^{ij}]
\right)
\right|^2\, .
\end{align}
Therefore 
\begin{align}
{\cal F}(\epsilon,0)&\sim \epsilon^{-\nu} \nonumber \\ \nu &= 2 - A[\sigma^{ij}].
\end{align}
Let us presume existence of a relation between the space-time cutoff $\epsilon$ and a regularization of the measure over the  field coupling to the source, ${\cal O}=\int d^d x \sqrt{\hat g} \sigma^{ij}T_{ij}$.  If this regularization can be interpreted as a course graining, or finite resolution, $\delta$ in the embedding space of ${\cal O}$, then comparing weights of ${\cal O}$ and $\epsilon$ under Weyl transformation suggests a $\delta \sim \epsilon^2$.  Hence,   
\begin{align}\label{nupr}
{\cal F}(\epsilon,0)&\sim \delta^{-\nu'} \nonumber\\
\nu' &= 1-\frac{A[\sigma^{ij}]}{2}
\end{align}
and  the analogue of the Lipshitz-H\"older exponent is 
\begin{align}
\alpha = 1-\nu' = \frac{A[\sigma^{ij}]}{2}\, ,
\end{align}
for the case $A[\sigma^{ij}]<2$, otherwise ${\cal F}(\epsilon,0)$ is finite as $\epsilon\rightarrow 0$.
For $d=2$, the integrated anomaly is purely topological, proportional to the Euler characteristic $\chi$;
\begin{align}
A = -\frac{c}{24\pi}\int\sqrt{g}R = -\frac{c}{6}\chi\, ,
\end{align}
where $c$ is the central charge.

Presuming divergence of ${\cal F}(\epsilon,0)$ as $\epsilon\rightarrow 0$, one defines the analogue of  fractal dimension by obtaining the minimum positive $\gamma$ such that 
\begin{align}\label{garf}
{\cal F}(\epsilon,\gamma)\equiv \int &dJ \left|Z_\epsilon(J)^2 J^{-\gamma}\right|
\end{align}
converges as $\epsilon\rightarrow 0$. AdS/CFT duality gives
\begin{align}
{\cal F}(\epsilon,\gamma) &\approx\int d{\cal G}\epsilon^{-2 +  A[ \sigma^{ij} ] + 2\gamma}
\left|
\exp\left(
-\tilde S_{cl}( {\cal G}\sigma^{ij})
\right)
\right|^2 \left|{\cal G}\right|^{-\gamma}\\
&\sim \epsilon^{-2+ A[\sigma^{ij}] + 2\gamma}\, ,
\end{align}
such that 
\begin{align}\label{gamsol}
\gamma = 1-\frac{A[ \sigma^{ij} ]}{2}\, .
\end{align}
In light of \eqref{nupr}, the bound $\nu'  \le \gamma$ is saturated, consistent with a fractal interpretation.  The fractal dimension is
\begin{align}
D_f = 1-\gamma = \frac{A[ \sigma^{ij} ]}{2}\,.
\end{align}
For AdS${}_{3}/$CFT${}_{2}$,
\begin{align}\label{thedim}
D_f = -\frac{c}{12}\chi = \frac{c}{6}(g-1)\, ,
\end{align}
where $g$ is the genus. 

The expression \eqref{thedim} yields a fractal dimension, provided the constraint  $0\le D_f \le 1$ is satisfied.  For sufficently large central charge or genus, ${\cal F}(\epsilon,0)$ is finite as $\epsilon\rightarrow 0$ and the dimension of the measure is the same as that of the embedding space, $D_f = D=1$.   A physical interpretation for the negative $D_f$ arising for genus zero, the Reimann sphere,  is unclear.   This  particular case is unique in that there is no choice of $\gamma$ for which  ${\cal F}(0,\gamma) = \int dJ|Z^2J^{-\gamma}|$ is finite if the integration contour includes a boundary at $J=0$.  Although $\gamma=(1-D_f)$  renders the integration convergent at large $J$, the integral diverges at $J=0$ since $\gamma>1$.   

In all the arguments above, analyticity in $J$ has been assumed such that AdS/CFT duality can be extended to arbitrarily large complex sources. While the assumption may appear brazen, the predicted large $J$ asymptotics can be verified explicitly in two dimensional conformal field theory, as shown below.

\section{Large source asymptotics in two-dimensions}\label{lsrc}

Consider a two dimensional conformal field theory with central charge $c$, fields $m$, metric $g$ and action $S[m,g]$.
The effective gravitational action $\Gamma[g]$ is defined by integrating out the matter fields;
\begin{align}
e^{-\Gamma[g]} = \int{\cal D}m\, e^{-S[m,g]}\, .
\end{align}
The term in the effective action giving rise to the Weyl anomaly is nonlocal \cite{Duff2}. However
by a suitable coordinate transformation one can write the metric $g_{ij}$ in the conformal gauge,
\begin{align}\label{gaug}
g_{ij}(x) = h_{ij} (x) e^{2\phi(x)}\,
\end{align}
for some fixed choice of the metric $h_{ij}$.
The effective action then has the  form \cite{Poly};
\begin{align}\label{nliou}
\Gamma[g] =  \tilde\Gamma[h]
+ \frac{c}{48\pi}\int d^2 x \sqrt{h}
\left(\frac{1}{2}h^{\mu\nu}\partial_\mu\phi \partial_{\nu}\phi +  \phi  R\right)
\end{align} 
where the second term, containing all information about the Weyl anomaly, is now local. 

Suppose there is a source corresponding to 
a deformation of the inverse metric,
\begin{align}\label{defg}
g^{ij}(x) \rightarrow  g^{ij}(x) + J\sigma^{ij}(x) \, .
\end{align}
At large $J$, \eqref{gaug} implies $\phi(\vec x) \approx -\frac{1}{2}\ln(J)$  
giving 
\begin{align}
Z(J)=e^{-\Gamma(g)} \sim J^{\frac{c}{96 \pi}\int d^2 x \sqrt{\tilde g}\tilde R} = J^{\frac{c}{24}\chi}\, .
\end{align}
Consequently, the minimum choice of $\gamma$ such that the integral  
\begin{align}
{\cal F}(0,\gamma)=\int dJ |Z(J)^2 J^{-\gamma}|
\end{align} 
converges at large $J$ is 
\begin{align}
\gamma = 1+\frac{c}{12}\chi\, ,
\end{align}
consistent with the result \eqref{gamsol} of the previous section.
Note that it is important in these arguments that \eqref{nliou} is not the Liouville action, lacking a cosmological constant term $\mu e^{2\phi}$ which would alter the asymptotic behavior.


\section{Information dimension and entropy}\label{sec5}
 
 The relation between the Weyl anomaly and fractal dimension resembles known results relating the Weyl anomaly to terms in the entanglement entropy which are logarithmically divergent as $\epsilon\rightarrow 0$  \cite{CWreplica,Holzhey,Calabrese,Fursaev,Soludukhin,Ryu2,Odds,CasiniFree,Dowker,MyersSinha,Hung,deBoer}.
 One definition of fractal dimension, known as the information dimension,  is the coefficient of a logarithmically divergent term in the Shannon entropy,
\begin{align}
D_{\rm information} = \lim_{\delta\rightarrow 0} \frac{-\sum_i P_i \ln(P_i)}{\ln(\delta)} \, ,
\end{align}
 where the fractal is covered with boxes of length $\delta$ in an embedding space  $\{X\}$,  and the  ``probability'' of being found in the box $i$ is $P_i \equiv \int_i d\mu(X)$.  
 It is often the case that inequivalent definitions of fractal dimension yield the same result, so that the a dimension determined from large $J$ asymptotics could be equivalent to the information dimension.
 Assuming this to be the case, the arguments in the previous section suggest that the log divergence of the Shannon entropy for a measure over a field ${\cal O}$ in a conformal field theory  is 
proportional to the Weyl anomaly.

Unlike the Shannon entropy, the entanglement entropy has divergences which are worse than logarithmic for theories in dimension greater than two, although these are non-universal, depending on the renormalization scheme. 
For an even dimensional conformal field theory,  the entanglement entropy across a surface $\Sigma$ has the form,
\begin{align}\label{uni}
S_{\rm entanglement}(\Sigma)= S_{\rm nonuniversal} + a\ln(\epsilon) + {\rm finite}.
\end{align}
where the universal term $a$ is obtained from the Weyl anomaly, and has been argued to satisfy a c-theorem, behaving monotonically under renormalization group flow \cite{Latorre,CasiniHuerta1,SolodRicci,KlebRG,Kleb3d,MyersArbDim,CasiniHuerta2,Casini3,Soludatheorem}. 
The information dimension of the path integral measure, or coefficient of the logarithmic divergence of the Shannon entropy, may be quite similar to the logarithmic divergence of the entanglement entropy, in that  both are derived from the Weyl anomaly.  It seem physically reasonable that the information dimension should decrease monotonically under renormalization group flow.

For odd dimensional space-times with no Weyl anomaly, there is no logarithmic divergence in the entanglement entropy.  The only universal term in this case is finite, and has been proposed as a candidate quantity for a c-theorem \cite{CasiniHuerta2,MyersArbDim,Casini3,Kleb3d}.  
In the case of the Shannon entropy and information dimension discussed above, we would seem to be presented with a similar conundrum regarding odd dimensions.  One would apparently conclude, based on the absence of a Weyl anomaly, that the information dimension is integer for odd space-time dimension.  

\section{Conclusions} 
We have argued that the Weyl anomaly in conformal field theory is  a close cousin of fractal dimensionality.  
The argument is based on a computation of  the large source asymptotics of the generating functional in conformal field theories admitting a holographic dual.  
Subject to certain assumptions regarding analytic continuation to arbitrarily large complex sources, the relation between the Weyl anomaly and the fractal dimension is more than just an analogy.  The analogues of the Lipschitz-H\"older exponent and fractal dimension computed using holographic duality are equal,  saturating an inequality known to hold for fractals.  
Just as the Weyl anomaly relates to logarithmic ultraviolet divergences in entanglement entropy, it is apparently also related to similiar divergences in the Shannon entropy of the path integral measure.  The  coefficient of the latter divergence is the fractal dimension known as the  information dimension.
The notion that c-theorems can be related to monotonic behavior of information dimension under renormalization group flow is intriguing. 
Equally interesting is the possibility that some known fractals may have a holographic representation in terms of a gravity-like theory.


  
\pagebreak


\end{document}